\documentclass[twocolumn,letter,10pt]{article}
\usepackage{procIAGssymp}
\usepackage{epsfig}
\usepackage{epsf}
\newcommand{\beq}{\begin{equation}}
\newcommand{\eeqn}{\nonumber\end{equation}}
\newcommand{\dss}{\displaystyle}
\newcommand{\eeq}[1]{\label{#1}\end{equation}}
\newcommand{\Frac}[2]{\frac{\displaystyle\strut #1}{\displaystyle\strut #2} }
\newcommand{\ifm}{\makebox{}\ifmmode}
\long\def\ifundefined#1#2#3{\expandafter\ifx\csname
     #1\endcsname\relax#2\else#3\fi}
\newcount\switch
\newcommand{\Begmat}{\ifm\switch=1\else\switch=0$\fi}
\newcommand{\Endmat}{\ifnum\switch=0$\fi}
\renewcommand{\deg}[1]{\Begmat #1^\circ\Endmat}

\newcommand{\bigint}{ {\mbox{\LARGE{$\dss\int$}}}}
\newcommand{\Gr}[1]{#1}
\newcommand{\Bl}[1]{#1}

\newcommand{\hr}{\rule[-1.0ex]{0.0em}{3.0ex}}

\newcommand{\nntab}[2]{ \multicolumn{2}{#1}{#2} }

\newcommand{\nnnntab}[2]{ \multicolumn{4}{#1}{#2} }

\title{Using a Numerical Weather Model to Improve Geodesy}
\author{Arthur Niell
      \thanks{MIT Haystack Observatory, Westford, MA, USA 01886
              (aniell@haystack.mit.edu)}
      \and Leonid Petrov
      \thanks{NVI, Inc./NASA Goddard Space Flight Center, Greenbelt,
              MD, USA (Leonid.Petrov@gsfc.nasa.gov)}
}
\date{ }
\begin{document}
\maketitle

\paragraph{Abstract.}

    The use of a Numerical Weather Model (NWM) to provide in situ atmosphere
information for mapping functions of atmosphere delay has been evaluated
using Very Long Baseline Interferometry (VLBI) data spanning eleven
years. Parameters required by the IMF mapping functions (Niell 2000, 2001)
have been calculated from the NWM of the National Centers for Environmental
Prediction (NCEP) and incorporated in the CALC/SOLVE VLBI data
analysis program. Compared with the use of the NMF mapping functions (Niell
1996) the application of IMF for global solutions demonstrates that the
hydrostatic mapping function, IMFh, provides both significant improvement in
baseline length repeatability and noticeable reduction in the amplitude of 
the residual harmonic site position variations at semidiurnal to long-period 
bands. For baseline length repeatability the reduction in the observed mean 
square deviations achieves 80\% of the maximum that is expected for the
change from NMF to IMF. On the other hand, the use of the wet mapping 
function, IMFw, as implemented using the NCEP data, results in a slight 
degradation of baseline length repeatability, probably due to the large grid 
spacing of the NWM that is used.

\section{Introduction}

    The accuracy of the estimate of the local vertical coordinate from GPS
and VLBI observations increases as the minimum elevation cutoff of included
data is reduced, due to improved decorrelation of the estimates of 
troposphere path delay and the site vertical coordinate. From the other side, 
the errors in mapping the zenith troposphere delay increase rapidly with 
decreasing elevation angle and cause a degradation of the precision of the 
estimate of site position coordinates. Thus, it is important to assess the 
optimum minimum elevation above which data should be included.

   The total atmosphere path delay is expressed as the product of the zenith
path delay and the mapping function, $m(e)$, defined as the ratio of
the excess path delay at geometric elevation, $e$, i.e., the vacuum
elevation, to the path delay in the zenith direction. The hydrostatic and wet
mapping functions are calculated separately  because the vertical distributions
of refractivity differ (MacMillan and Ma, 1994; Niell, 1996). Many geodetic
analyses for GPS and VLBI currently use very simple models for mapping
functions that are based on site location and time of year (Niell, 1996;
hereafter referred to as NMF). This climatological approach is used due
to the difficulty of obtaining in situ meteorological information along the
path traversed by the incoming radio waves.

   The next best thing would be mapping functions based on the vertical
distribution of refractivity above the site, assuming azimuthal symmetry.
While measurement of such a profile, for example by launching radiosondes
several times per day, has not been deemed feasible, some amount of
information on the state of the atmosphere can be obtained from the
meteorologic numerical weather models (NWM), which provide estimates of the
vertical distribution of temperature, pressure, and water vapor content at
specified horizontal grid points over the entire globe every six hours.
Niell (2000, 2001) demonstrated how this information can be used
to improve the mapping functions.

   In this note we show that these mapping functions (IMF), using data
from NCEP on a $2.\!^\circ5$~by~$2.\!^\circ5$  grid interpolated to the site
location, provide the expected improvement in baseline length repeatability,
and that, furthermore, they provide a noticeable reduction in the amplitudes
of residual harmonic site position variations.

   While the wet mapping function (IMFw) has the potential to be an
improvement over NMFw, as demonstrated using radiosonde profiles (Niell,
2000), this is not achieved in the current implementation using the NCEP NWM
in the VLBI analysis package CALC/SOLVE (Ma et al., 1990).

   In section 2 we give a brief description of IMF and in section 3 outline 
the implementation in CALC/SOLVE. The VLBI analysis is described in section 4 
and some of the geodetic implications are presented in sections 5. Section 6 
contains a summary of the results and suggestions for future developments.

\section{The Isobaric Mapping Functions (IMF)}

   Currently the closest approximation to a "true" mapping function is obtained 
by using radiosonde data to define the vertical profile of radio refractivity 
and, assuming spherical symmetry about the radiosonde launch site, integrating 
along the path determined by the refractivity to determine the delay through 
the atmosphere both in the zenith direction and at the elevation for which 
the mapping function is to be calculated. The delays due to the hydrostatic 
and wet components of the atmosphere (Niell 1996) are retained independently 
in the raytrace integration, and the mapping functions are calculated 
separately.

   Marini (1972) showed that, for a spherically symmetric but vertically
arbitrary profile of atmospheric refractivity, the mapping function can be
approximated by a continued fraction of the form
\beq
     m(e) = \Frac{1 + \Frac{a}{ 1 + \Frac{b}{1+c} } }
                 {\sin{e} + \Frac{a}{ \sin{e} + \Frac{b}{\sin{e}+c} } }
\eeq{e:e1}
The numerator is included to normalize the fraction at an elevation angle
of \deg{90}, and the number of terms (only three are shown) should be
determined by the desired accuracy of the fit. Niell (1996) found that
three terms is sufficient to keep the error to less than 1 mm for
elevations down to \deg{3} if the delays are calculated at eight elevations
in addition to the zenith direction.

   The goal of mapping function research is to find the functional form
and dependence on external information of the parameters $a$, $b$,
$c$  \ldots\ that best matches the actual values of the mapping function for
real atmospheres at arbitrary locations and times.

   In order to provide a useful tool for estimating or calculating the
atmosphere path delay for geodetic observations, the parameters $a$, $b$,
and $c$ should be calculable in terms of available information. The
following procedure was used to determine the relation of the parameters
to external data.

\begin{enumerate}
    \item For radiosonde profiles from a large number of sites and spanning at
          least one year, calculate, by raytracing, the hydrostatic and wet
          mapping functions at nine elevations from 90 degrees down to 3 
          degrees.

    \item Calculate by least-squares fit the continued fraction parameters $a$,
          $b$, and $c$ for each profile for both hydrostatic and wet
          mapping functions.

    \item Assume that the parameters are linear functions of some other
          parameters, $q_i$, that are available to characterize the mapping
          function, such as geographic location, time of year, or a 
          combination of meteorological parameters. Expand each of the 
          parameters $a$, $b$, and $c$ in terms of the $q_i$.

    \item Use linear regression of all $a$ values on parameters $q_i$ for all
          sites and profiles to determine the regression coefficients that
          describe $a$ in terms of the $q_i$. Repeat for $b$ and $c$; do
          for both the hydrostatic and wet parameters.
\end{enumerate}

   To determine the parameter dependence, the hydrostatic and wet
mapping functions were calculated from radiosonde profiles for twenty-six
sites for a one year period (1992) with usually two profiles per day, at 00
and 12 UT. Based on the expectation that the hydrostatic mapping function 
would be strongly correlated with the thickness of the atmosphere, a possible
empirical relation to the geopotential heights of constant pressure levels
(isobaric surfaces) for each of the parameters $a_h$, $b_h$, and $c_h$ was
investigated. The 200 hPa level gave the best agreement (Niell, 2000).
The relation was found to require a dependence on the cosine of twice the
latitude. The coefficients relating the parameters $a_h$, $b_h$, and
$c_h$ to the geopotential height and the latitude were obtained by linear
regression on the 200 hPa heights from the radiosonde data and
{\em cos(2*latitude)}. Since the 200 hPa isobar has little sensitivity to the
topography below, a correction for height of a site above sea level was
calculated by linear regression of the site heights and the residual mapping
function error using the derived coefficients. This was found to be consistent
with that used for NMF, and the NMF height correction (Niell 1996) was adopted.

   Although the concept of atmospheric thickness failed to produce a 
reasonable relation for the wet mapping function, we found that there exists 
a linear regression between the parameters $a_w$, $b_w$, and $c_w$ and a ''wet
pseudo-mapping function''. The wet pseudo-mapping function, $\rho$, is given by
\par\noindent\vspace{-2ex}
\beq
      \rho = \Frac{r(3^\circ\!\!.3)}{r(90^\circ)}
\eeq{e:i1}
\par\noindent\vspace{-3ex}
where
\par\noindent\vspace{-2ex}
\beq
      r(\epsilon) = \Frac{ k_3 \bigint \Frac{e_v(h)}{T^2(h)}
           \Frac{1 + \frac{h}{R_\oplus}}{\sqrt{
           \left(1+\frac{h}{R_\oplus}\right)^2
         - \cos^2 \epsilon }} \, dh}{k_3 \int \Frac{e_v(h)}{T^2(h)} \, dh}
\eeq{e:i2}
$\rho$ is the ratio of the integral of the wet refractivity along
an elevation of \deg{3.3} to the integral in the vertical direction, and
$r(\epsilon)$ is the integral of wet refractivity along elevation angle
$\epsilon$. Here $e_v$ is specific humidity, $T$ is temperature, $h$ is 
height, and the integration is done through the whole atmosphere. 
$r(\epsilon)$ is a pseudo-mapping function, because bending of the raypath 
due to the spatial change in the index of refraction is not included in the 
calculation. This parameter was investigated in order to be able to avoid 
the large amount of computing required for an accurate ray-trace.

   These regression coefficients allows us to compute mapping functions
using geopotential height and the ratio of the wet refractivity integrals
obtained from a numerical weather model (NWM) for the case when no radiosonde
atmosphere profile near an observing station is available, as is usually
the case during VLBI and GPS observations.

\section{Implementation at GSFC for VLBI}

   The NWM data used are the gridded data from the U.S. National Center for
Environmental Protection (NCEP) that are output every six hours. We used
the Reanalysis model (Kalnay et al., 1996) in our study. It appears on their
Web site with a time lag of 3 to 7 days. 
All Reanalysis model products have been analyzed with a consistent, though 
not necessarily the most recent, model. The current model was adopted in 1996.
The files that are downloaded contain geopotential height and temperature at
eighteen pressure levels, specific humidity at eight pressure levels, and the
surface height for each grid point. The values of geopotential height at 200
hPa are extracted, and the values of $\rho$ are calculated at each grid point.
For the VLBI application the parameters are subsequently interpolated to the
horizontal position of each VLBI site at each NCEP epoch (0, 6, 12, 18 UT).
During interpolation, the unit vector of the normal to the surface of
geopotential height at 200 hPa is computed. We assume  that the hydrostatic
atmosphere path delay is azimuthally symmetric with respect to the normal
to the geopotential height, rather than local zenith. Thus, we imply that the
atmosphere is tilted with respect a local observer. Typical values of the tilt
are 0.1 to 1.0 mrad.

\section{VLBI analysis and results}

   The troposphere path delay at each station is modeled as a linear spline
with typical time span of 20 minutes, and parameters of the linear spline
are estimated together with other parameters of interest in a single
least-squares solution. The hydrostatic mapping function is used for computing
the apriori hydrostatic path delay (the apriori wet path delay is set
to zero) and the wet mapping function is used as the partial derivative
for estimation of the wet zenith path delay. If the hydrostatic mapping
function is wrong, then the apriori total path delay in the direction of the 
observed source will be biased. Since the hydrostatic mapping function and the 
wet mapping function are slightly different at low elevations (the difference
is 4--7\% at elevation \deg{5}), the estimation procedure will not remove this
bias completely. Thus, systematic errors in the hydrostatic mapping function
result in systematic errors in estimates of site position. Due to the errors
in the wet mapping function, which is used as the partial derivative, the
contribution of the wet troposphere path delay to the observables is not
completely removed by estimating the wet path delay in the zenith direction.

\begin{figure}[!h]
     \centerline{\epsfclipon \epsfxsize=74mm \epsffile{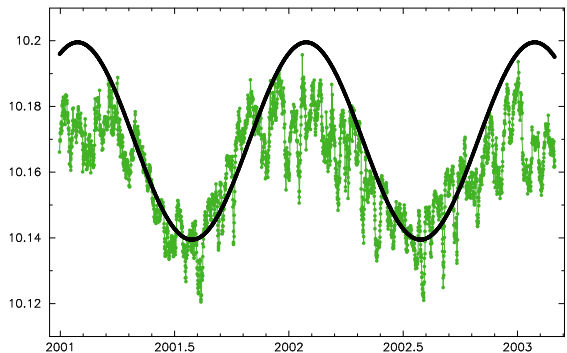} }
     \caption{ NMFh (thick line) and IMFh (dots) at elevation angle $5^\circ$
              at station Gilcreek (Alaska).}
     \label{f:p1}
\end{figure}

\begin{figure}[!h]
     \centerline{\epsfclipon \epsfxsize=74mm \epsffile{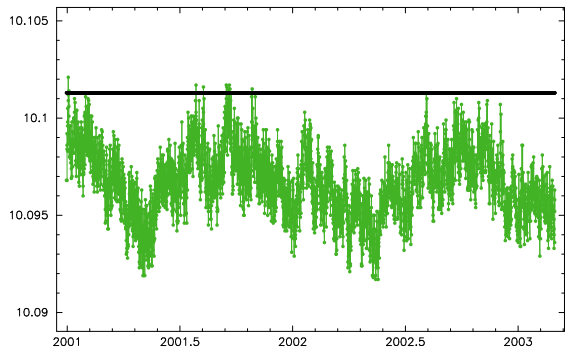} }
     \caption{NMFh (thick line) and IMFh (dots) at elevation angle $5^\circ$
              at station Fortleza (Atlantic coast, Brazil).}
     \label{f:p2}
\end{figure}

   Figures \ref{f:p1} and \ref{f:p2} show the differences between the
hydrostatic NMF and IMF mapping functions at two sites. If the IMF is closer
to the true mapping function, we could expect that using IMF would reduce
systematic errors in GPS and VLBI results.

   To evaluate {\em quantitatively} what kind of improvement in geodetic
results is provided by the use of IMF instead of NMF, we performed a series
of VLBI solutions and investigated changes in baseline length repeatability
and changes in the estimated amplitudes of the harmonic site position
variations.

\subsection{Baseline length repeatability test}

\begin{figure}[!h]
    \centerline{\epsfclipon \epsfxsize=74mm \epsffile{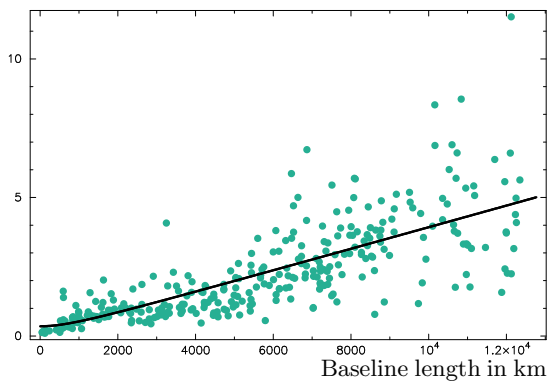} }
    \par\vspace{-1ex}\par\hspace{44mm}{\small Baseline length in km}
    \caption{Changes in baseline lengths due to replacement of NMFh with
             IMFh (in mm) as a function of baseline length.}
     \label{f:p3}
\end{figure}

   First, we need to evaluate how a replacement of the NMFh with IMFh affects
baseline lengths. For this purpose we made a special solution. We used the
standard parameterization in the estimation model: we solved for site
positions, troposphere path delay modeled by spline of the first degree with
time span 20 minutes, clocks, Earth orientation parameters, source positions,
and other nuisance parameters. The hydrostatic zenith path delay was computed
using the formula of Saastamoinen (1972), and IMFh was used to
transform the apriori zenith hydrostatic path delay to the path delay in the
source direction. However, in the right hand-side of the equations of conditions,
instead of the differences between observed and modeled delay we inserted the 
differences between hydrostatic path delays calculated using NMFh and 
IMFh but with the same apriori zenith hydrostatic path delay computed from the 
Saastamoinen formula. All VLBI observations for the period of 1993--2003 were
used. We obtained from this solution the series of
baseline length estimates which have the meaning of changes in baseline 
lengths due to changes in the hydrostatic mapping function, and we computed the
weighted mean value of the length change for each baseline. Since replacement 
of the mapping function affects mainly the estimate of vertical site 
coordinate, the changes in baseline length, $\delta l$, will depend on 
baseline length as
$ \Delta l = \Frac{L}{2 R_\oplus} \, (\Delta h_1 +  \Delta h_2) $, where
$L$ is the baseline length and $R_\oplus$ is the radius of the Earth.
Figure \ref{f:p3} shows the mean changes in baseline length as a function
of the total baseline length for all baselines in the solution.

\begin{figure}[!h]
    \centerline{\epsfclipon \epsfxsize=74mm \epsffile{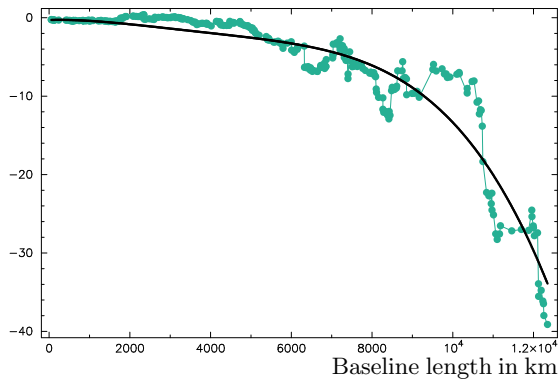} }
    \par\vspace{-1ex}\par\hspace{44mm}{\small Baseline length in km}
    \caption{Differences in squares of baseline lengths in $\mbox{mm}^2$
             using IMFh versus NMFh for computing apriori hydrostatic path 
             delay, in the sense IMFh-NMFh.}
     \label{f:p4}
\end{figure}

   We then made two other solutions. In solution A we used NMFh and in
solution B we used IMFh. Parameterization was exactly the same as in the
previous special solution. For each baseline we got the series of length
estimates and evaluated the formal uncertainties of these estimates.
A linear model was fitted to each series, and the wrms deviation from
the linear model was computed for each baseline. We call this wrms baseline
length repeatability. We formed the differences of the squares of
repeatabilities in the sense solution A (IMFh) minus solution B (NMFh).
The differences, smoothed with a Gaussian filter, are presented as a function
of baseline length in figure~\ref{f:p4}.

   We see that the differences are negative. This means that using IMFh instead
of NMFh reduced the baseline length repeatability and, thus, improved the 
quality of the geodetic results. But we would like to have a {\em quantitative 
measure} of how replacement of NMFh by IMFh brings us closer to the true 
mapping function. We introduce a coefficient of reduction of variance $R$
(see Petrov and Boy (2003) for more details) as a quantitative measure of
changes in baseline length repeatability:

\beq
       R = \Frac{\Delta \sigma^2 + \sigma_m^2}{2 \, \sigma_m^2}
\eeq{e:a1}
where $\Delta \sigma^2$ is the difference in the squares of baseline length
repeatability of the solution with NMFh minus the solution with IMFh,
and $\sigma_m^2$ is the square of differences in baseline lengths, derived
from the special solution, which has the meaning of the square of expected
changes in baseline lengths. This coefficient runs from 0 to 1. In the case
that the change of mapping function did not result in a change of baseline
length repeatability, the coefficient is 0.5. In the case that NMFh is much
closer to the (unknown to us) "true" mapping function than IMFh is, the use of
IMFh will result in degradation of baseline length repeatability by the amount
$\sigma_m^2$, and then $R \approx 0$. If IMFh is closer to the true mapping
function than NMFh by an amount equal to the difference (NMFh - IMFh), then the
difference in baseline length repeatability is equal to $\sigma_m^2$, and
$R=1$.

   We have computed the coefficient of reduction of variance for each baseline
observed 16 or more times and have calculated the weighted mean value of
$R$ over all baselines. We found that $ R = 0.81 \pm 0.02 $. This means that
80\% of the power in the change of baseline lengths due to the change in
mapping function is present in the data, and the baseline length repeatability
has improved by approximately 80\% of the expected change under the assumption
that the new mapping function, IMFh, is closer to the true mapping function
than the old NMFh by the amount of the differences between these two functions.

   This result is very encouraging. It demonstrates that using only one
{\em value}, the geopotential height of the 200 hPa surface, we can
reconstruct a mapping {\em function} (an infinite set of values) with
a surprisingly good precision: not worse than 80\%.

\subsection{Changes in harmonic site position variations}

   As we see in figures~\ref{f:p1} and \ref{f:p2}, NMFh does not model
seasonal changes perfectly, and it completely ignores harmonic variations 
of the mapping function at frequencies others than annual. Thus, we can expect 
that these errors will cause harmonic variations in the estimates of site 
positions. As we saw in the previous section, the use of IMFh improves baseline
length repeatability. Therefore we can expect that observed residual harmonic 
site position variations will be reduced when IMFh is used instead of NMFh.

   In order to evaluate this effect quantitatively, we made two
solutions. In both solutions we estimated troposphere path delay, clock
parameters, Earth orientation parameters, station position and velocities,
source coordinates, and harmonic site position variations at 32 frequencies
for each station. NMFh/NMFw were used in the first solution, and IMFh/NMFw 
were used in the second solution. We included all observations from 1980 
through 2003 for 40 VLBI stations with a long history of observations. Our
theoretical model included site displacements due to solid Earth tides, ocean
loading, atmosphere loading, and hydrology loading. Thus, we obtained residual
harmonic site position variations.

   For each frequency we computed the ratio of the weighted sum of squares of
the observed amplitudes of harmonic site position variations to its
mathematical expectation, $\chi^2/f$. In the absence of a signal $\chi^2/f$ 
should be less than 1.2. Detailed analysis of the technique for solutions of 
this type can be found in Petrov and Ma (2003).

   We noticed differences in these statistics at six frequencies. They are
presented in table~\ref{t:t1}. We see that the power of the residual signal
is reduced at all frequencies when IMFh is used instead of NMFh. To our
surprise the power was reduced not only at annual (Sa), diurnal ($S_1$) and
semi-diurnal ($S_2$) frequencies, where we expected improvement, but also
at semi-annual (Ssa), sidereal ($K_1$) and semi-sidereal ($K_2$) frequencies.
Changes of the power of residual signal at the $K_1$ frequency may have
a significant consequence: it means that using IMFh instead of NMFh may result
in a non-negligible change in the estimate of the precession rate, since 
precession corresponds to the harmonic variation in the Earth rotation at the 
sidereal frequency.

\begin{table}[!h]
\caption{ $\chi^2/f$ statistics of residual harmonic site position variations
            in two solutions which used different mapping functions.}
\begin{center}
     \begin{tabular}{|l | l @{\hspace{1em}} l  | l @{\hspace{1em}} l|}
         \hline
         Tide & \nnnntab{c|}{ $\chi^2/\mbox{f}$ }                         \hr \\
         \hline
              & \nntab{|c|}{\Bl{ NMFh, NMFw}} & \nntab{c|}{\Gr {IMFh}, NMFw}
                                                                          \hr \\
              & \quad vert       & horiz  & \qquad vert   & horiz         \hr \\
         \hline
         $ \sf K_2 $   & \quad  \Bl{ 1.80 } & \Bl{ 2.01 } &
                         \qquad \Gr{ 1.61 } & \Gr{ 2.05 }     \hr \\
         $ \sf S_2 $   & \quad  \Bl{ 2.55 } & \Bl{ 1.73 } &
                         \qquad \Gr{ 2.15 } & \Gr{ 1.76 }     \hr \\
         $ \sf K_1 $   & \quad  \Bl{ 2.13 } & \Bl{ 3.45 } &
                         \qquad \Gr{ 1.89 } & \Gr{ 3.34 }     \hr \\
         $ \sf S_1 $   & \quad  \Bl{ 3.80 } & \Bl{ 2.32 } &
                         \qquad \Gr{ 3.58 } & \Gr{ 2.22 }     \hr \\
           \sf Ssa     & \quad  \Bl{ 2.42 } & \Bl{ 1.12 } &
                         \qquad \Gr{ 1.85 } & \Gr{ 1.09 }     \hr \\
           \sf Sa      & \quad  \Bl{ 6.34 } & \Bl{ 2.56 } &
                         \qquad \Gr{ 5.77 } & \Gr{ 2.43 }     \hr \\
     \hline
     \end{tabular}
\end{center}
\label{t:t1}
\vspace{-3ex}
\end{table}

\subsection{Changes in the scale factor of the TRF}

   Systematic changes in the estimates of the vertical component of site
positions may result in a change of the scale factor of the output site 
position catalogue. In order to evaluate quantitatively this effect, we 
have analyzed the catalogue of site positions obtained in the solutions 
described in the previous section. We estimated the 7-parameter transformation 
of the site position catalogue from the solution that used IMFh with respect
to the catalogue from the solution that used NMFh. This transformation
includes translation, rotation, and the scale factor. We found that for the
site position catalogue from solutions with IMFh the scale is greater
by $ 0.21 \pm 0.05 $ ppb with respect to the solutions using NMFh. This changes 
in the scale factor corresponds to an approximately 1 mm increase of vertical 
coordinates of site positions.

\subsection{Wet IMF}

   We made the baseline length test for using IMFw in place of NMFw, but
instead of improvement in the baseline length repeatability, we found a slight
degradation. It should be noted that the expected improvement, 1.0 mm at
baseline lengths 10000 km, is very small, a factor of 5 less than the
improvement due to replacement of the hydrostatic mapping function. At the
same time comparison of IMFw and NMFw with respect to the mapping function
derived using radiosonde data at the sites of radiosonde launches indicates
that IMFw better approximates the radiosonde mapping function than NMFw,
in contradiction to the results of the VLBI data analysis. We think that this
contradiction is related to the fact that the relatively large grid of the
NCEP Reanalysis NWM, 200x270 km at mid-latitudes, is not adequate to represent
the wet refractivity field.

\section{Geodetic implications}

   If there were no model errors, it would be advantageous to make
observations to as low an elevation as permitted by the equipment and
the physical surrounding, for both GPS and VLBI. Including the widest range
of elevations reduces the uncertainty in the height estimate and improves
separation of the estimates of the vertical component of position from the
estimates of the zenith path delays and the clock offsets. However, any error
in the azimuthally symmetric atmosphere model will propagate into
errors in the estimated vertical coordinate, and these errors will increase 
with decreasing minimum elevation. Thus, there is a trade-off between 
atmosphere uncertainty and the accuracy of determination of the vertical, 
and it is possible to choose the optimum minimum elevation if the error model 
is sufficiently well known.

   The formal uncertainty in the vertical, taking into account atmosphere
stochastic variability, assuming the clock is modeled as white noise, and
adding a fixed amount of delay noise, is calculated in the precise point 
positioning mode (PPP) of the GPS analysis software Gipsy (Zumberge et al., 
1997), although other effects, such as satellite orbit error, are not included 
explicitly. For twenty-four hour sessions the uncertainties for minimum 
elevations of \deg{15} down to \deg{5} decrease from 4.0 mm to 2.0 mm (the 
data were taken with an Ashtech Z-12 with good sky visibility).

\begin{figure}[!h]
     \centerline{ \epsfig{file=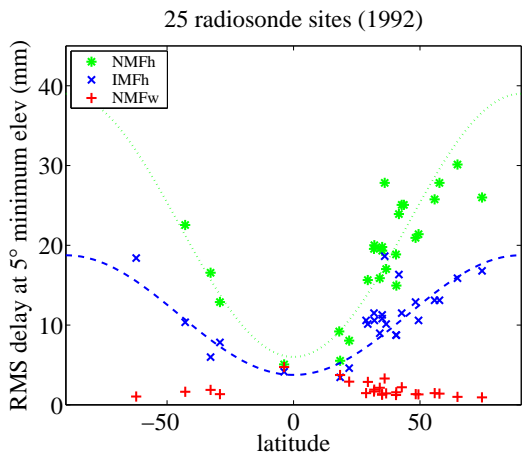,width=2.75in} }
     \protect\caption{Rms delay error for mapping functions at radiosonde sites
                 for $5^\circ$ elevation.}
     \label{f:b}
\end{figure}

   The effect of the atmosphere model error has been calculated by comparing
the IMF and NMF mapping functions obtained from the NWM at twenty-six
radiosonde launch sites for the year 1992, with the mapping functions
calculated by raytracing along the path of the radio waves at various
elevations, assuming a vertical distribution of refractivity given by the
radiosonde temperature, pressure, and humidity profiles. Using the surface
pressure for the zenith hydrostatic delay, and the zenith wet delay from the
radiosonde data, the rms delay errors for an elevation of \deg{5} are shown
in figure~\ref{f:b} for the hydrostatic component of IMF and for the 
hydrostatic and wet components of NMF. The errors for IMFw are very close 
to those of NMFw.

   How do these delay errors affect the height estimates? For errors that 
increase with decreasing elevation, such as those that are due to the 
atmosphere, the lowest elevation observations have the greatest effect on the 
height uncertainty. As a result, the effect on the height uncertainty can be 
calculated as the product of the delay error at the lowest observed elevation 
and the sensitivity of the height estimate to delay change at that elevation.

    MacMillan and Ma (1994) evaluated the sensitivity of the vertical 
error to the error in atmosphere path delay at the minimum elevation.
They found that an error of 1 mm in atmosphere path delay at the minimum 
elevation angle causes 0.22 mm error in the vertical coordinate if the minimum 
elevation is \deg{5}, and 0.4 mm if the minimum elevation is \deg{10}. We made 
a similar test for a later set of VLBI data and found a sensitivity of 
0.3--0.35 for a minimum elevation of \deg{5}. A simulation of GPS observations 
using a simple model of estimating vertical, zenith troposphere delay, 
and clock for a single epoch gave sensitivity of 0.4 to 0.8 over the minimum 
elevation range \deg{5}--\deg{15}. For GPS the actual sensitivity will depend 
on what satellites are visible at the time of observation, but for 
illustration of the technique we will use the values from our GPS simulation.

\begin{figure}[!h]
     \centerline{ \epsfig{file=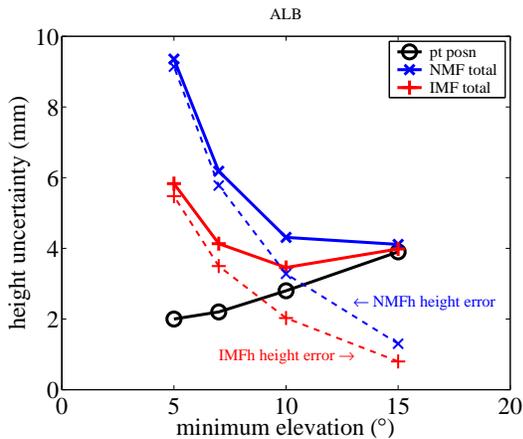,width=2.75in}  }
     \caption{Height uncertainty for different minimum elevations for
              precise point-position and for mapping functions.}
     \label{f:c}
\end{figure}

   To show the effect of the mapping function errors on height at 
mid-latitude the rms delay differences were found between both the NMF and 
IMF mapping functions and mapping functions obtained by raytracing the 
radiosonde data (as described in section 2) for the site ALB (Albany, state 
New York, USA) at latitude \deg{42}. The differences were converted to 
height uncertainties using the sensitivities described in the previous 
paragraph (figure~\ref{f:c}). We calculated inflated formal uncertainties 
of the height estimates by adding in quadrature the point-positioning 
uncertainties and the uncertainty due to errors in the mapping function. 
These total errors are also shown in figure~\ref{f:c}. From this 
study we conclude that the use of NMFh significantly increases the total 
height uncertainty when data are included below about \deg{10} for a 
twenty-four hour session. At the same time the use of IMFh does not result 
in an increase of total height uncertainties until data below about \deg{7} 
are included.

   This type of evaluation can be important for deciding how much data
to include in a solution. For example, the lower elevation data might
be desirable for improving estimates of atmosphere gradients, but only
if the uncertainty of vertical estimates is not degraded.

   An important point to notice from figure~\ref{f:c} is that for a minimum
elevation of \deg{15}, which is the standard for many GPS analyses, the error
in the vertical estimation due to the atmosphere is less than 2 mm and
is therefore not significant for daily solutions. However, in computing weekly
site position averages, the atmosphere modeling error may contribute more than
expected since weather systems often persist over many days, and the actual
uncertainty of the average will decrease more slowly than $\frac{1}{\sqrt{n}}$,
because the errors are correlated over longer than one day.

\section{Summary, conclusions, future work}

   Atmosphere delay mapping functions, designated IMF, based on in situ
meteorological information from the NCEP numerical weather model, have been
implemented in the VLBI analysis package CALC/SOLVE. When applied to global
VLBI solutions that include all data above a minimum elevation of \deg{5},
their use, compared to the model NMF, improves baseline length repeatability
and reduces the power in residual harmonic site position variations.

   If used for GPS, IMF will allow lower elevation data (down to 
approximately \deg{7}) to be included in the solutions while maintaining the 
formal uncertainty in the vertical that is obtained for \deg{15} minimum 
elevation, and will reduce 
amplitudes of observed residual harmonic site position variations at annual 
and semi-annual periods.

   With improvement in modeling of annual and semi-annual atmosphere errors,
it is time to begin modeling systematic temperature-dependent antenna height 
errors for GPS and VLBI.

\section{Acknowledgment}

   We used NCEP Reanalysis data provided by the NOAA-CIRES Climate
Diagnostics Center, Boulder, Colorado, USA, from their Web site at
{\mbox\tt http://www.cdc.noaa.gov/}

\section*{References}
\parindent=0pt
\def \cita{\par \hangindent=4mm \hangafter=1}

\cita Kalnay~E. et al.,
       (1996). The NCEP/NCAR 40-Year Reanalysis
       Project, {\it Bullet. Amer. Meteorol. Soc.}, 77, p.437--471.

\cita Ma,~C., J.M.~Sauber, L.J.~Bell, T.A.~Clark, D.~Gordon, W.E.~Himwich,
      and J.W.~Ryan, (1990). Measurement of Horizontal Motions in Alaska
      Using Very Long Baseline Interferometry,
      {\it J. Geophys. Res.}, 95(B13), p.~21,991--22,011.

\cita MacMillan,~D.~S., and C.~Ma, (1994). Evaluation of very long baseline
      interferometry atmospheric modeling improvements, {\it J. Geophys.
      Res.}, 99(B1), p.~637-651.

\cita Marini,~J.~W., (1972). Correction of satellite tracking data for
      an arbitrary tropospheric profile, {\it Radio Science}, 7, p.~223--231.

\cita Niell, A.E., (1996). Global mapping functions for the atmosphere delay
      at radio wavelengths, {\it J. Geophys. Res.}, 101, No. B2,
      p.~3227--3246.

\cita Niell,~A.~E., (2000). Improved atmospheric mapping functions for VLBI
      and GPS, {\it Earth, Planets, and Space}, 52, p.~699--702.

\cita Niell,~A.~E., (2001). Preliminary evaluation of atmospheric mapping
      functions based on numerical weather models, {\it Phys. Chem. Earth},
      26, p.~475--480.

\cita Petrov,~L., and C.~Ma, (2003). Study of harmonic site position variations
      determined by VLBI, {\it J. Geophys. Res.},  vol. 108, No. B4, p.~2190,
      doi: 10.1029/2002JB001801.

\cita Petrov~L., and J.-P.~Boy, (2003). Study of the atmospheric pressure
      loading signal in VLBI observations, submitted to {\it J. Geophys. 
      Res.}.

\cita Saastamoinen,~J. (1972). Atmospheric correction for the troposphere
      and stratosphere in radio ranging of satellites. In {\it The Use of
      Artificial Satellites for Geodesy, Geophys. Monogr.}, AGU, 15,
      p.~247--251.

\cita Zumberge,~J.~F., M.~B.~Heflin, D.~C.~Jefferson, M.~M.~Watkins,
      and F.~H.~Webb, (1997).  Point positioning for the efficient and robust
      analysis of GPS data from large networks. {\it J. Geophys. Res.},
      102, p.~5005--5017.

\end{document}
\endinput